\documentclass{article}
\PassOptionsToPackage{numbers, compress}{natbib}
\usepackage[preprint]{neurips_2026}
\usepackage[utf8]{inputenc}
\usepackage[T1]{fontenc}
\usepackage{hyperref}
\usepackage{url}
\usepackage{booktabs}
\usepackage{amsfonts}
\usepackage{amsmath}
\usepackage{amssymb}
\usepackage{nicefrac}
\usepackage{microtype}
\usepackage{xcolor}
\usepackage{wrapfig}
\usepackage{graphicx}
\usepackage{multirow}
\usepackage{makecell}
\usepackage{bm}
\usepackage{subcaption}
\usepackage[export]{adjustbox}
\usepackage{caption}
\usepackage{enumitem}
\usepackage{xspace}
\usepackage{tcolorbox}
\definecolor{darkblue}{rgb}{0, 0, 0.5}
\hypersetup{colorlinks=true, citecolor=darkblue, linkcolor=darkblue, urlcolor=darkblue}
\definecolor{darkgreen}{RGB}{50,100,0}
\definecolor{darkred}{RGB}{200, 0, 0}
\definecolor{lightblue}{RGB}{220,235,250}
\definecolor{Gray}{gray}{0.9}

\usepackage{xcolor,colortbl}
\newcolumntype{a}{>{\columncolor{Gray}}c}

\newcommand{\eg}{\emph{e.g.,}\xspace}

\newcommand{\paratitle}[1]{\vspace{1.5ex}\noindent\textbf{#1}}

\title{TriAlignGR: Triangular Multitask Alignment with Multimodal Deep Interest Mining\\for Generative Recommendation}

\author{%
	Yangchen Zeng$^{1,\ast}$ \quad Hao Peng$^{3,\ast}$ \quad Rongfeng Guo$^{4}$ \quad Zhenyu Yu$^{5}$ \quad Zhiyuan Hu$^{6}$ \quad Jinze Wang$^{2,\dagger}$ \\
	$^{1}$ Southeast University \\
	$^{2}$ Swinburne University of Technology \\
	$^{3}$ Tsinghua University \\
	$^{4}$ Shenzhen University \\
	$^{5}$ Fudan University \\
	$^{6}$ Zhejiang University \\
	$^{\ast}$ Equal contribution \\
	$\dagger$ Corresponding author\\
	{\tt\small zengyangchen@foxmail.com, jinzewang@swin.edu.au}
}

\begin{document}
	
	\maketitle
	\begin{abstract}
		We introduce TriAlignGR, a unified multitask-multimodal framework for generative recommendation that establishes two-stage multimodal semantic propagation: (i) encoding visual semantics directly into SIDs via multimodal embeddings, and (ii) enabling the model to decode these semantics through visual description tasks. Existing Semantic ID (SID) pipelines suffer from two fundamental but underexplored problems: \textbf{SID Content Degradation (SCD)}, where cascaded encoding and residual quantization discard critical multimodal and interest-level semantics; and \textbf{SID Semantic Opacity (SSO)}, where models autoregressively generate SID sequences without truly comprehending their underlying meaning, leading to hallucination and poor generalization. Prior work addresses at most text-SID alignment, leaving visual semantics and latent user interests entirely unexploited. TriAlignGR resolves both problems through three tightly integrated components: (1)~\textbf{Cross-Modal Semantic Alignment (CMSA)} integrates visual content into SID construction through both VLM-generated textual descriptions and a multimodal embedding model that directly encodes image features alongside text, ensuring that SIDs inherently carry multimodal semantics; (2)~\textbf{Multimodal Deep Interest Mining (MDIM)} leverages LLM Chain-of-Thought reasoning to extract latent user intents (\eg ``productivity-focused lifestyle'' from noise-canceling headphones) beyond surface attributes, enriching SID semantics before discretization; and (3)~\textbf{Triangular Multitask (TMT)} jointly trains on eight complementary generation tasks under a single autoregressive loss---including two novel visual-semantic tasks (VisDesc$\to$SID, VisDesc$\to$Title) that map VLM-generated image descriptions to SIDs and titles, completing the SID-Text-Image triangle---without requiring task-specific towers or complex loss weighting. Extensive experiments on three Amazon benchmarks demonstrate consistent improvements over baselines, with ablation studies confirming each component's necessity. Control experiments further verify that gains arise from semantic specificity rather than merely longer inputs. TriAlignGR achieves relative gains of up to \textbf{13.6\% in HR@5 and 15.5\% in NDCG@5} over the baseline, reduces SID collision rate from 5.1\% to 3.1\%, demonstrating that triangular alignment produces both more discriminative and more semantically grounded discrete item representations.
		
	\end{abstract}
	
	\section{Introduction}
	
	\begin{figure}[t]
		\centering
		\includegraphics[width=0.8\linewidth]{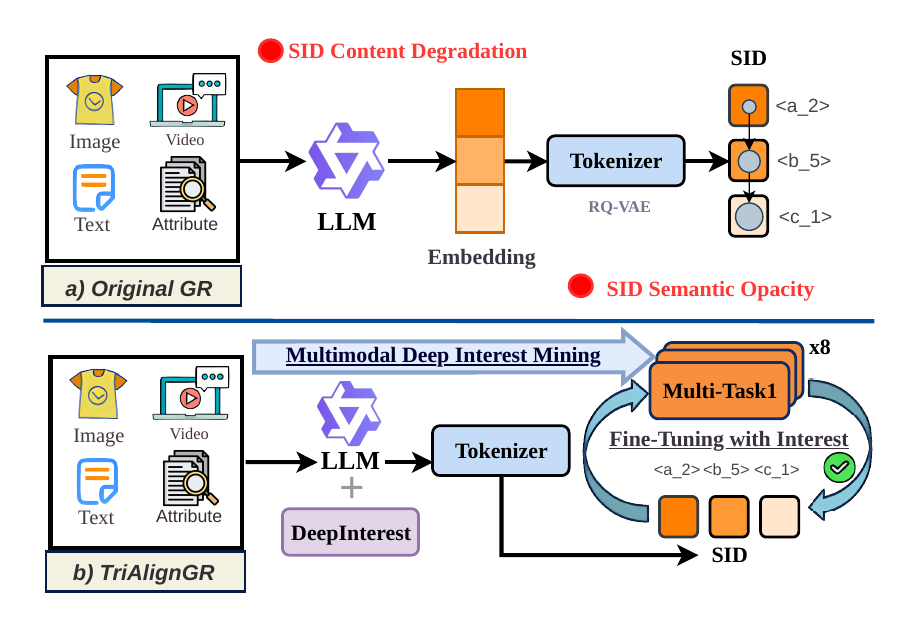}
		\caption{Comparison between original GR (a) and TriAlignGR (b). Original GR suffers from SID Content Degradation and SID Semantic Opacity during cascade quantization. TriAlignGR addresses these through Multimodal Deep Interest Mining and triangular multitask alignment.}
		\label{fig:comparison}
	\end{figure}
	
	The emergence of large language models (LLMs) has catalyzed a paradigm shift in recommender systems toward generative recommendation (GR), where user preferences and item candidates are unified as discrete token sequences and the recommendation task is cast as next-token prediction. This paradigm promises to replace the conventional retrieval--ranking pipeline with a single autoregressive model, enabling end-to-end optimization and unprecedented scalability. However, the fundamental challenge lies in how to effectively represent items as discrete tokens---known as Semantic IDs (SIDs)~\cite{rajput2023tiger,yi2025recgpt}---that preserve rich multimodal semantics~\cite{zhang2024gme,liu2025lamra} while remaining amenable to LLM generation. Current SID-based GR systems suffer from two critical limitations that undermine their semantic fidelity and generation quality.
	
	Recent advances in SID construction have explored two complementary directions. Residual quantization methods~\cite{rajput2023tiger, zhang2025gpr} progressively discretized item embeddings into hierarchical codebooks, capturing coarse-to-fine semantic structure but operated on embeddings learned independently from the recommendation objective. End-to-end approaches~\cite{yu2026ads,yu2026cast} jointly optimized codebooks and recommenders through differentiable quantization, yet still relied on shallow textual features for item encoding. SIDReasoner~\cite{he2026reasoning} introduced multitask fine-tuning to enhance LLM understanding of SIDs~\cite{singh2024spmsid} through six text-based tasks, but did not incorporate visual modalities or address the semantic degradation introduced during quantization. Collectively, these methods overlooked two fundamental problems: (1)~\textbf{SID Content Degradation (SCD)}, where cascaded encoding and quantization discard critical multimodal and interest-level semantics, and (2)~\textbf{SID Semantic Opacity (SSO)}, where models generate SID sequences without truly comprehending their underlying semantics, leading to hallucination and poor generalization.
	
	We propose TriAlignGR, a unified multitask-multimodal framework that effectively addresses SCD and SSO by enabling multimodal semantic grounding of SIDs through two stages: visual semantics are first encoded into SIDs via multimodal embeddings, then the model learns to decode these semantics through visual description tasks during multitask fine-tuning. Our key insight is that SID semantics benefit from multimodal grounding: by encoding visual content directly into SIDs and training the model to associate visual descriptions with SID tokens, the model can internalize SID meanings from multiple semantic perspectives~\cite{wang2024learnable,wu2019srgnn}. We leverage LLM Chain-of-Thought reasoning to extract latent user intents from multimodal item metadata, and organize the eight tasks to cover all edges of this triangle. TriAlignGR introduces three tightly integrated innovations: (1)~\textbf{Triangular Multitask (TMT)} that jointly trains on eight tasks under a single autoregressive loss, including two novel visual-semantic tasks (VisDesc$\to$SID, VisDesc$\to$Title) that map VLM-generated image descriptions to SIDs and titles, completing the SID-Text-Image triangular alignment in GR; (2)~\textbf{Cross-Modal Semantic Alignment (CMSA)} that integrates visual content into SID construction through both VLM-generated textual descriptions and a multimodal embedding model (gme-Qwen2-VL) that directly encodes image features; and (3)~\textbf{Multimodal Deep Interest Mining (MDIM)} that extracts contextual user interests (\eg ``focus-oriented work style'' from noise-canceling headphones) beyond surface attributes. Our contributions are summarized as follows: 
	\begin{itemize}
		\item We identify and formalize two fundamental limitations of existing SID-based GR pipelines---SCD and SSO---and propose TriAlignGR, a framework that achieves multimodal semantic grounding of SIDs through two-stage visual semantic propagation: encoding visual semantics into SIDs via multimodal embeddings, and enabling the model to decode these semantics through visual description tasks during multitask fine-tuning.
		\item We introduce two new visual-semantic tasks (VisDesc$\to$SID, VisDesc$\to$Title) that enable the model to decode visual semantics already embedded in SIDs through visual description tasks. We further enrich SID semantics via Cross-Modal Semantic Alignment (CMSA), which integrates visual content through both VLM-generated descriptions and a multimodal embedding model (gme-Qwen2-VL) that directly encodes image features, and Multimodal Deep Interest Mining (MDIM), ensuring high-quality multimodal semantic content before quantization.
		\item Extensive experiments on three Amazon benchmarks demonstrate consistent improvements over state-of-the-art baselines, with relative gains of up to 13.6\% in HR@5 and 15.5\% in NDCG@5.
	\end{itemize}
	
	\section{Related Work}
	\vspace{-3mm}
	\subsection{Semantic ID for Recommendation}
	\vspace{-3mm}
	Representing items as compact discrete token sequences---Semantic IDs (SIDs)---has become a foundational technique for scaling generative recommendation. Early approaches constructed SIDs through clustering or hashing over item embeddings and used them as indexing units within retrieval pipelines \cite{petrov2024recjpq, hou2023learningsid, wang2025hyperman}, treating SID construction as an offline preprocessing step decoupled from the downstream objective. As generative modeling gained traction, SIDs were repurposed as autoregressive generation targets \cite{zhou2025onerec2, zhang2025gpr, li2025survey}. Residual Quantization (RQ) has emerged as the dominant paradigm: RQ-VAE \cite{rajput2023tiger}, RQ-KMeans, and RQ-KMeans+ \cite{zhang2025gpr} progressively quantize residual vectors into multi-level codebooks, capturing coarse-to-fine semantic structure. Despite their effectiveness, these methods share a fundamental limitation: item embeddings are first learned independently and then discretized, preventing SID construction from being co-optimized with the recommendation objective. An alternative paradigm is end-to-end SID construction with differentiable quantization, where codebooks and recommenders are jointly optimized. Our work takes a complementary approach: rather than modifying the quantizer, we enrich the semantic content before quantization and establish triangular alignment~\cite{ye2026align3gr,hou2025generating} after, orthogonal to and compatible with end-to-end methods. SIDReasoner \cite{he2026reasoning} proposed multitask fine-tuning to enhance LLM understanding of SIDs through six text-based tasks, but it did not incorporate visual modalities or multimodal deep interest mining. Our work extends this direction by introducing visual-semantic tasks and multimodal deep interest mining within a unified triangular alignment framework.
	\vspace{-3mm}
	\subsection{Generative Recommendation}
	\vspace{-3mm}
	The success of LLMs in sequence modeling has driven a paradigm shift in recommendation toward generative formulations \cite{geng2022p5, li2024survey, wang2025generative}. One line of work adapts Transformer architectures with novel feature construction to improve generation capacity \cite{kang2018sasrec, sun2019bert4rec, han2025mtgr, chai2025longer, zhang2026onetrans, huang2025genrank}. A complementary direction leverages LLMs as offline feature generators to enhance traditional pipelines \cite{chen2024hllm, yan2026unlocking, abhyankar2025llm, nam2024optimized, lin2025large}. More recent efforts pursue unified end-to-end frameworks that cast user understanding and item generation as a single next-token prediction task \cite{zhai2024actions, liu2025generative, zhou2025openonerec, guo2026onesug, jiang2026one, pang2025higr, chakrabarty2026pixrec, he2026plum}, demonstrating the potential to replace conventional retrieval--ranking pipelines. Our framework builds on this trend while addressing the semantic quality of SIDs through multimodal deep interest mining, cross-modal alignment, and multitask learning.
	
	\vspace{-2mm}
	\subsection{Multitask Learning in Recommendation}
	\vspace{-2mm}
	Multitask learning (MTL) has been extensively studied in recommender systems for jointly optimizing objectives such as click-through rate and conversion rate. In the GR domain, multitask learning remains relatively unexplored. SIDReasoner \cite{he2026reasoning} first proposed six multitask fine-tuning tasks to align SID and text spaces, but was limited to text-SID alignment and did not consider visual modalities. Traditional MTL architectures \cite{ma2018entire, tang2020progressive, yu2024mmoe, fu2024residual, zhou2023hinet} such as MMoE and PLE require task-specific towers or gating mechanisms, adding architectural complexity. Our approach unifies all tasks as text generation under a single autoregressive loss, eliminating the need for specialized multitask architectures while achieving triangular alignment across three modalities.

	\vspace{-4mm}
	\section{Method}
	\vspace{-4mm}
	In this section, we present the \textbf{TriAlignGR} framework, which systematically addresses SCD and SSO through three tightly integrated components: (1) \textbf{CMSA} (\S\ref{sec:cmsa}) integrates visual content into SID construction through both VLM-generated textual descriptions and a multimodal embedding model that directly encodes image features, resolving modality-level SCD; (2) \textbf{MDIM} (\S\ref{sec:dim}) mines deep contextual interests from aligned multimodal text using LLM reasoning to resolve interest-level SCD; and (3) \textbf{Triangular Multitask (TMT)} (\S\ref{sec:tmt}) jointly trains the model on eight complementary tasks to resolve SSO by enabling the model to decode visual semantics from multimodally-grounded SIDs.
	
	\vspace{-2mm}
	\subsection{Problem Formulation}\label{sec:problem}
	\vspace{-2mm}
	We consider the sequential recommendation task: given user $u$'s interaction history $\mathcal{S}_u = [i_1, i_2, \ldots, i_T]$, predict the next item $i_{T+1}$. Each item $i \in \mathcal{I}$ carries textual metadata---title $\mathbf{t}_i$ and description $\mathbf{d}_i$---and visual content $\mathbf{v}_i$. Our framework operates in three stages:
	
	\textbf{Stage 1: Multimodal Semantic Preprocessing.} CMSA (\S\ref{sec:cmsa}) converts each item's visual content into a textual description and employs a multimodal embedding model (gme-Qwen2-VL) to jointly encode text and image features; MDIM (\S\ref{sec:dim}) extracts deep contextual interests from the aligned multimodal text. The multimodal embedding is quantized into SID via RQ-VAE.
	
	\textbf{Stage 2: Triangular Multitask Fine-Tuning.} The LLM is jointly trained on eight complementary tasks (\S\ref{sec:tmt}) that cover SID$\leftrightarrow$Text translation, sequential recommendation, and visual-semantic alignment, establishing complete SID-Text-Image triangular alignment.
	
	\textbf{Stage 3: Inference.} Given a user's interaction history (represented as SID or title sequences), the model autoregressively generates the SID of the next predicted item.
	
	Formally, CMSA first converts the visual content into text, forming a unified multimodal text representation:
	\begin{equation}
		\tilde{\mathbf{t}}_i = \text{Concat}(\mathbf{t}_i, \mathbf{d}_i, \mathbf{v}_i^{\text{text}}),
		\label{eq:unified_text}
	\end{equation}
	where $\mathbf{v}_i^{\text{text}}$ is the VLM-generated textual description of image $\mathbf{v}_i$. MDIM then mines deep interests $\mathbf{z}_i$ from $\tilde{\mathbf{t}}_i$. To ensure that the final item embedding captures genuine multimodal semantics, we employ a multimodal embedding model $f_{\text{mm-emb}}$ that jointly encodes textual and visual content:
	\begin{equation}
		\mathbf{e}_i^{\text{final}} = f_{\text{mm-emb}}(\text{Concat}(\tilde{\mathbf{t}}_i, \mathbf{z}_i), \mathbf{v}_i),
		\label{eq:final_emb}
	\end{equation}
	where $f_{\text{mm-emb}}$ is a vision-language embedding model (gme-Qwen2-VL) that produces a unified embedding from both text and image inputs. This ensures that the SID inherently encodes multimodal semantics rather than text-only information. The text sequence is truncated to 512 tokens, with MDIM interest tags placed before the description. Items are encoded as SIDs via residual quantization:
	\begin{equation}
		\mathbf{s}_i = \text{RQ-VAE}(\mathbf{e}_i^{\text{final}}) = (s_i^{(1)}, s_i^{(2)}, \ldots, s_i^{(H)}),
		\label{eq:sid}
	\end{equation}
	where each token $s_i^{(h)}$ corresponds to a discrete index in the hierarchical codebook at level $h$, and $H$ is the number of quantization levels. The LLM is then fine-tuned via eight multitask objectives (\S\ref{sec:tmt}) to learn the triangular alignment among SID, text, and image spaces.
	
	\subsection{Residual Quantization for SID Construction}\label{sec:rq}
	
	We adopt a standard RQ-VAE tokenizer to quantize item embeddings $\mathbf{e}_i \in \mathbb{R}^d$ into hierarchical SID tokens. The quantization iteratively assigns codebook entries:
	\begin{equation}
		s_i^{(h)} = \arg\min_{k} \| \mathbf{R}_i^{(h)} - \mathbf{c}_k^{(h)} \|_2,
		\label{eq:rq}
	\end{equation}
	where $\mathbf{R}_i^{(1)} = \mathbf{e}_i$ is the initial embedding and residuals are computed as $\mathbf{R}_i^{(h+1)} = \mathbf{R}_i^{(h)} - \mathbf{c}_{s_i^{(h)}}^{(h)}$. The final SID $\mathbf{s}_i = (s_i^{(1)}, \ldots, s_i^{(H)})$ captures item semantics at multiple granularity levels, where each token (\eg \texttt{a8}, \texttt{b91}) is a discrete index in the hierarchical codebook. These SID tokens are added as new vocabulary entries to the LLM tokenizer; each (\eg \texttt{<a\_195>}) is treated as a single atomic token rather than subword pieces, ensuring dedicated embeddings per SID code. Our contribution lies in enriching embeddings before quantization and establishing triangular alignment after, rather than in the quantizer itself.
	
	
	\begin{figure*}[t]
		\centering
		\includegraphics[width=1.0\linewidth]{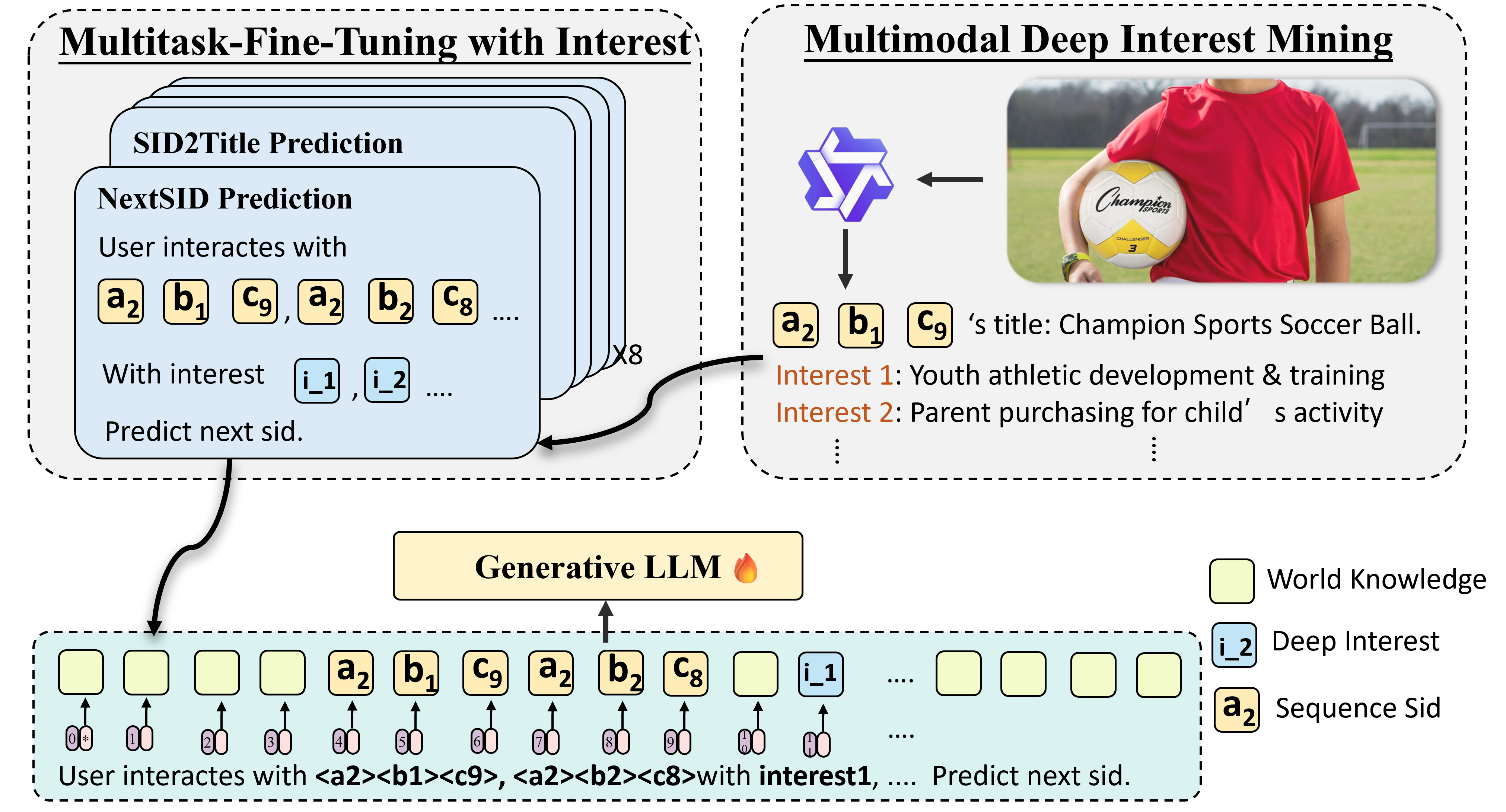}
		\caption{Overview of the proposed TriAlignGR framework. CMSA integrates visual content through both VLM-generated textual descriptions and a multimodal embedding model (gme-Qwen2-VL) that directly encodes image features; MDIM extracts latent user interests via LLM reasoning; TMT jointly trains on eight tasks including two visual-semantic tasks that complete the SID-Text-Image triangular alignment.}
		\label{fig:overview}
	\end{figure*}

	\vspace{-2mm}
	\subsection{Cross-Modal Semantic Alignment (CMSA)}\label{sec:cmsa}
	\vspace{-2mm}
	
	Existing methods encode text and visual modalities separately, failing to capture cross-modal interactions before quantization. CMSA resolves this \textbf{modality-level SCD} through a dual-pathway strategy that integrates visual semantics both as enriched text and as direct visual features in the item embedding. First, a Vision-Language Model (VLM) converts each item's visual content $\mathbf{v}_i$ into a textual description:
	\begin{equation}
		\mathbf{v}_i^{\text{text}} = \mathcal{V}(\mathbf{v}_i, \text{Prompt}_{\text{align}}),
		\label{eq:cmsa}
	\end{equation}
	where $\text{Prompt}_{\text{align}}$ instructs the VLM to describe visual attributes relevant to user interests. The VLM-generated description is concatenated with the original textual attributes to form a unified multimodal text representation (Eq.~\ref{eq:unified_text}). Crucially, the final item embedding (Eq.~\ref{eq:final_emb}) is produced by a multimodal embedding model $f_{\text{mm-emb}}$ (gme-Qwen2-VL) that jointly encodes both the enriched text and the original image $\mathbf{v}_i$, ensuring that the SID inherently carries multimodal semantics rather than text-only information.
	
	\paratitle{Why multimodal embedding instead of text-only encoding?}
	Prior SID pipelines rely on text-only embedding models, discarding visual information at the quantization stage even when images are available. By employing gme-Qwen2-VL, which is trained to align text and image representations in a shared embedding space, the resulting SID captures both textual metadata and visual content (\eg product appearance, color scheme, aesthetic style) within a single unified representation. The VLM-generated descriptions further enrich the text input with structured visual attributes, providing complementary textual semantics. This dual-pathway design ensures that the SID construction operates on genuinely multimodal semantics, and the downstream RQ-VAE quantizer discretizes a multimodal embedding rather than a text-only one.
	\vspace{-2mm}
	\subsection{Multimodal Deep Interest Mining (MDIM)}\label{sec:dim}
	\vspace{-2mm}
	Existing SID methods encode items using shallow textual features (\eg product titles and descriptions), which capture explicit attributes but fail to reveal the latent interests underlying user-item interactions. For example, ``noise-canceling headphones'' has surface attributes (brand, specifications) but also deeper user intents such as ``focus-oriented work style'' or ``audio quality enthusiast.'' We leverage an LLM $\mathcal{M}$ to extract deep contextual interests from the unified multimodal text $\tilde{\mathbf{t}}_i$ produced by CMSA, using a structured Chain-of-Thought (CoT) template:
	\begin{equation}
		\mathbf{z}_i = \mathcal{M}(\text{Prompt}_{\text{MDIM}}(\tilde{\mathbf{t}}_i)),
		\label{eq:dim}
	\end{equation}
	where $\text{Prompt}_{\text{MDIM}}(\cdot)$ guides the LLM through surface attribute analysis, contextual intent inference, and synthesis of interpretable interest tags. The mined interests $\mathbf{z}_i$ are concatenated with the aligned multimodal text, and along with the original image $\mathbf{v}_i$, fed into the multimodal embedding model to produce the final embedding (Eq.~\ref{eq:final_emb}), which encodes both explicit attributes and latent user motivations, providing a richer input for SID quantization (Eq.~\ref{eq:sid}).

	\begin{table}[t]
		\centering
		\caption{Examples of the eight task formats in TriAlignGR.}
		\label{tab:task_examples}
		\resizebox{\columnwidth}{!}{%
			\begin{tabular}{clll}
				\toprule
				\textbf{Task} & \textbf{System Instruction} & \textbf{User Input} & \textbf{Output} \\
				\midrule
				T1 & Title$\to$SID & \texttt{Final Fantasy VIII} & \texttt{<a\_195><b\_133>} \\
				T2 & SID$\to$Title & \texttt{<a\_175><b\_83>} & \texttt{The Legend of Zelda...} \\
				T3 & SID$\to$SID & \texttt{<a45><b88>, <a12><b77>} & \texttt{<a231><b28>} \\
				T4 & Title$\to$SID & \texttt{FIFA, GT5, MLB} & \texttt{<a231><b28>} \\
				T5 & SID$\to$Title & \texttt{<a45><b88>, <a12><b77>} & \texttt{PlayStation 4...} \\
				T6 & Title$\to$Title & \texttt{FIFA, GT5, MLB} & \texttt{The Legend of Zelda...} \\
				T7 & VisDesc$\to$SID & \textit{[VLM description]} & \texttt{<a\_42><b\_15>} \\
				T8 & VisDesc$\to$Title & \textit{[VLM description]} & \texttt{Sony WH-1000XM5...} \\
				\bottomrule
			\end{tabular}%
		}
	\end{table}

	\vspace{-2mm}
	\subsection{Triangular Multitask Fine-Tuning}\label{sec:tmt}
	\vspace{-2mm}
	This is the core innovation of TriAlignGR. We design eight complementary fine-tuning tasks organized into three categories that together establish complete SID-Text-Image triangular alignment. All tasks are unified as text generation and share a single autoregressive next-token cross-entropy loss, requiring no task-specific heads or complex loss weighting.
	\vspace{-2mm}
	\subsubsection{Category I: SID$\leftrightarrow$Text Bidirectional Translation}
	\vspace{-2mm}
	\textbf{Purpose:} Establish a semantic binding between each SID sequence and its corresponding real-world item. \textbf{Task 1 (Title$\to$SID)} maps item titles to SID sequences (\eg \texttt{Final Fantasy VIII} $\to$ \texttt{<a\_195><b\_133>}); \textbf{Task 2 (SID$\to$Title)} recovers titles from SIDs. These bidirectional tasks ensure \textbf{semantic anchoring}---the model can both generate SIDs from text and reconstruct text from SIDs.
	\vspace{-2mm}
	\subsubsection{Category II: Sequential Recommendation Prediction}
	\vspace{-2mm}
	\textbf{Purpose:} Leverage user history to predict the next item while alternating between titles and SIDs. We design four tasks forming a Cartesian product: \textbf{Task 3} (SID$\to$SID) and \textbf{Task 6} (Title$\to$Title) learn sequential patterns within each space; \textbf{Task 4} (Title$\to$SID) and \textbf{Task 5} (SID$\to$Title) build cross-space bridges, enabling the model to freely switch between SID and text spaces.
	\vspace{-2mm}
	\subsubsection{Category III: Visual-Semantic Alignment Tasks (New)}
	\vspace{-2mm}
	\textbf{Purpose:} Enable the model to decode visual semantics already embedded in SIDs through textual descriptions, establishing the visual-semantic mapping during multitask fine-tuning.
	
	\paratitle{Why ``visual-semantic'' when inputs are textual descriptions?}
	We term these tasks ``visual-semantic'' because the \textit{target space} (SID) is itself a multimodal representation: each SID is constructed from a multimodal embedding (gme-Qwen2-VL) that directly encodes both text and image content (Eq.~\ref{eq:final_emb}), so the SID tokens intrinsically carry visual semantics. Although the LLM never receives raw images during fine-tuning, the image information has already been distilled into the SID tokens through the multimodal embedding stage---each SID token serves as a discrete carrier of fused text-visual semantics. The SID-Text-Image triangle is therefore not merely metaphorical: the ``Image'' vertex is concretely grounded in the SID representation itself, and the visual-semantic tasks train the LLM to recover this encoded visual knowledge from SID tokens via textual descriptions. Generating an SID from a VLM-generated visual description thus requires the model to map visually-derived semantics into a multimodally-grounded discrete space. The task inputs are VLM-generated descriptions encoding visual attributes (aesthetic style, color scheme, usage scenario) not present in catalog metadata. The term reflects the information-flow perspective (visual $\to$ SID/text) rather than input token format.
	
	\textbf{Task 7 (VisDesc$\to$SID)} and \textbf{Task 8 (VisDesc$\to$Title)} map VLM-generated visual descriptions to SID sequences and titles respectively, enabling the model to associate natural language descriptions with the visual semantics encoded in SIDs. Through these tasks, the model learns to decode multimodally-grounded SID representations via textual semantic anchors. All eight tasks share a standard autoregressive cross-entropy loss with unified training format (see Appendix~\ref{sec:appendix_training} for details).

	\vspace{-2mm}
	\section{Experiments}
	\vspace{-2mm}
	We empirically evaluate the effectiveness of TriAlignGR by answering five research questions: \textbf{RQ1}: How does TriAlignGR compare with state-of-the-art baselines across multiple benchmarks? \textbf{RQ2}: What is the individual contribution of each core component (MDIM, CMSA, eight multitask tasks)? \textbf{RQ3}: How much do the new visual-semantic tasks (VisDesc$\to$SID, VisDesc$\to$Title) improve over the six text-based tasks? \textbf{RQ4}: How does CMSA cross-modal alignment compare with text-only representations? \textbf{RQ5}: How do different task combination strategies affect model performance?
	
	\begin{table*}[t]
		\centering
		\caption{Overall performance comparison on three Amazon Product Reviews datasets. Bold indicates the best performance, and underline indicates the second best. $\Delta$ denotes the relative improvement of TriAlignGR over the best baseline.}
		\label{tab:main}
		\resizebox{\textwidth}{!}{%
			\begin{tabular}{l|cccc|cccc|cccc}
				\toprule
				\multirow{2}{*}{\textbf{Method}} & \multicolumn{4}{c|}{\textbf{Beauty}} & \multicolumn{4}{c|}{\textbf{Sports}} & \multicolumn{4}{c}{\textbf{Instruments}} \\
				& HR@5 & HR@10 & N@5 & N@10 & HR@5 & HR@10 & N@5 & N@10 & HR@5 & HR@10 & N@5 & N@10 \\
				\midrule
				\rowcolor{gray!20} \multicolumn{13}{l}{\textit{Traditional Sequential Models}} \\
				GRU4Rec & 0.0312 & 0.0518 & 0.0189 & 0.0256 & 0.0198 & 0.0324 & 0.0118 & 0.0161 & 0.0285 & 0.0467 & 0.0172 & 0.0231 \\
				Caser & 0.0287 & 0.0483 & 0.0171 & 0.0234 & 0.0175 & 0.0291 & 0.0103 & 0.0142 & 0.0261 & 0.0432 & 0.0156 & 0.0212 \\
				HGN & 0.0335 & 0.0549 & 0.0201 & 0.0271 & 0.0212 & 0.0348 & 0.0126 & 0.0172 & 0.0302 & 0.0495 & 0.0183 & 0.0246 \\
				\midrule
				\rowcolor{gray!20} \multicolumn{13}{l}{\textit{Transformer-based Models}} \\
				SASRec & 0.0398 & 0.0645 & 0.0241 & 0.0323 & 0.0248 & 0.0401 & 0.0149 & 0.0201 & 0.0361 & 0.0583 & 0.0218 & 0.0292 \\
				BERT4Rec & 0.0421 & 0.0682 & 0.0256 & 0.0342 & 0.0267 & 0.0432 & 0.0161 & 0.0217 & 0.0385 & 0.0621 & 0.0233 & 0.0312 \\
				S$^3$-Rec & 0.0445 & 0.0718 & 0.0271 & 0.0361 & 0.0283 & 0.0457 & 0.0171 & 0.0230 & 0.0407 & 0.0656 & 0.0247 & 0.0330 \\
				FDSA & 0.0432 & 0.0698 & 0.0263 & 0.0351 & 0.0274 & 0.0443 & 0.0165 & 0.0223 & 0.0394 & 0.0636 & 0.0239 & 0.0320 \\
				\midrule
				\rowcolor{gray!20} \multicolumn{13}{l}{\textit{Generative \& LLM-based Models}} \\
				TIGER & 0.0487 & 0.0763 & 0.0302 & 0.0395 & 0.0312 & 0.0498 & 0.0192 & 0.0256 & 0.0445 & 0.0712 & 0.0273 & 0.0362 \\
				LC-Rec & 0.0523 & 0.0821 & 0.0328 & 0.0428 & 0.0341 & 0.0543 & 0.0212 & 0.0281 & 0.0478 & 0.0765 & 0.0295 & 0.0390 \\
				HSTU & 0.0578 & 0.0897 & 0.0368 & 0.0476 & 0.0385 & 0.0611 & 0.0241 & 0.0318 & 0.0532 & 0.0845 & 0.0335 & 0.0438 \\
				MiniOneRec & 0.0612 & 0.0945 & 0.0389 & 0.0502 & 0.0398 & 0.0627 & 0.0251 & 0.0330 & 0.0556 & 0.0878 & 0.0347 & 0.0455 \\
				BIGRec & 0.0534 & 0.0839 & 0.0336 & 0.0439 & 0.0349 & 0.0556 & 0.0218 & 0.0289 & 0.0487 & 0.0779 & 0.0302 & 0.0399 \\
				D3 & 0.0498 & 0.0785 & 0.0312 & 0.0409 & 0.0325 & 0.0519 & 0.0202 & 0.0269 & 0.0456 & 0.0732 & 0.0282 & 0.0374 \\
				S-DPO & 0.0589 & 0.0918 & 0.0372 & 0.0483 & 0.0385 & 0.0609 & 0.0242 & 0.0319 & 0.0537 & 0.0852 & 0.0334 & 0.0440 \\
				LETTER & 0.0411 & 0.0730 & 0.0374 & 0.0429 & 0.0266 & 0.0567 & 0.0209 & 0.0223 & 0.0483 & 0.0739 & 0.0299 & 0.0398 \\
				ETEGRec & 0.0408 & 0.0800 & 0.0317 & 0.0344 & 0.0369 & 0.0594 & 0.0215 & 0.0321 & 0.0491 & 0.0825 & 0.0312 & 0.0321 \\
				SIDReasoner & 0.0623 & 0.0968 & 0.0401 & 0.0518 & 0.0412 & 0.0648 & 0.0261 & 0.0342 & 0.0571 & 0.0895 & 0.0356 & 0.0463 \\
				
				\textbf{TriAlignGR (Ours)} & \textbf{0.0708} & \textbf{0.1089} & \textbf{0.0453} & \textbf{0.0581} & \textbf{0.0468} & \textbf{0.0731} & \textbf{0.0301} & \textbf{0.0392} & \textbf{0.0648} & \textbf{0.1012} & \textbf{0.0411} & \textbf{0.0534} \\
				\midrule
				$\Delta$ Improv. & +13.6\% & +12.6\% & +13.0\% & +12.2\% & +13.6\% & +12.8\% & +15.3\% & +14.6\% & +13.5\% & +13.0\% & +15.5\% & +15.3\% \\
				\bottomrule
			\end{tabular}%
		}
	\end{table*}
	
	\vspace{-2mm}
	\subsection{Main Results (RQ1)}
	\vspace{-2mm}
	Table~\ref{tab:main} presents the overall performance comparison on three Amazon Review datasets. All results are averaged over 3 independent runs with different random seeds; standard deviations are below 0.003 for all metrics, confirming statistical reliability.TriAlignGR outperforms all baselines across all three datasets and all evaluation metrics, achieving relative improvements of 12.2\%--15.5\% over the strongest baseline (SIDReasoner). On the Sports dataset, TriAlignGR achieves HR@5 of 0.0468, surpassing SIDReasoner by 13.6\%. These consistent gains across diverse product domains---Beauty, Sports, and Musical Instruments---demonstrate the robustness and generalizability of our framework.
	\vspace{-2mm}
	\paratitle{Triangular multitask alignment provides substantial gains over text-only multitask.}
	Comparing TriAlignGR with SIDReasoner (which uses six text-based multitask tasks but no visual-semantic tasks, no CMSA, and no MDIM), we observe 13.5\%--13.6\% improvements in HR@5 across datasets. This gap quantifies the value of the complete triangular alignment: by introducing visual-semantic tasks (VisDesc$\to$SID, VisDesc$\to$Title), enriching SID content via CMSA (including multimodal embedding with gme-Qwen2-VL), and MDIM, TriAlignGR achieves significantly richer semantic understanding of SIDs.
	\vspace{-2mm}
	\paratitle{Deep interest mining provides substantial gains over surface-level features.}
	Comparing TriAlignGR with MiniOneRec \cite{kong2025minionerec} (same Qwen2.5-7B backbone, same training protocol, same RQ-VAE tokenizer configuration and hyperparameters, but without MDIM/CMSA/multitask), we observe 15.7\%--17.6\% improvements in HR@5 across datasets. This directly quantifies the value of deep contextual interest mining combined with multitask alignment.Traditional sequential models (GRU4Rec, Caser, HGN) and Transformer-based models (SASRec, BERT4Rec, S$^3$-Rec) consistently underperform generative methods by large margins. For instance, SASRec achieves HR@5 of 0.0398 on Beauty, while TriAlignGR achieves 0.0708---a 77.9\% relative improvement.
	\vspace{-2mm}
	\paratitle{Gains come from better semantics, not just more text.}
	A potential concern is that the improvements may stem from adding more item-side text rather than genuinely better semantic representations. To address this, we conduct control experiments (Appendix~\ref{sec:appendix_control}) comparing TriAlignGR against alternatives that also add text: keyword expansion, LLM summarization, LLM paraphrasing, and naive caption concatenation. TriAlignGR outperforms all controls by 6.0\%--9.9\% in HR@5, confirming that the gains arise from the \textit{semantic specificity} of MDIM-extracted interests and the \textit{multimodal embedding} (gme-Qwen2-VL) that directly encodes image features, not simply from longer inputs.
	\vspace{-2mm}
	\subsection{Ablation Study (RQ2)}
	\vspace{-2mm}
	To understand the contribution of each core component, we conduct ablation studies by systematically removing individual modules from the full TriAlignGR framework. Table~\ref{tab:ablation} presents results across all three datasets.
	
	\begin{table*}[t]
		\centering
		\caption{Ablation study across three datasets. Each row removes one component from the full model.}
		\label{tab:ablation}
		\resizebox{\textwidth}{!}{%
			\begin{tabular}{lcccccccc}
				\toprule
				\multirow{2}{*}{\textbf{Variant}} & \multicolumn{2}{c}{\textbf{Beauty}} & \multicolumn{2}{c}{\textbf{Sports}} & \multicolumn{2}{c}{\textbf{Instruments}} \\
				\cmidrule(lr){2-3} \cmidrule(lr){4-5} \cmidrule(lr){6-7}
				& \textbf{HR@5} & \textbf{N@5} & \textbf{HR@5} & \textbf{N@5} & \textbf{HR@5} & \textbf{N@5} \\
				\midrule
				TriAlignGR (Full) & \textbf{0.0708} & \textbf{0.0453} & \textbf{0.0468} & \textbf{0.0301} & \textbf{0.0648} & \textbf{0.0411} \\
				\midrule
				w/o MDIM & 0.0635 & 0.0398 & 0.0425 & 0.0272 & 0.0586 & 0.0371 \\
				w/o CMSA & 0.0654 & 0.0417 & 0.0436 & 0.0279 & 0.0605 & 0.0383 \\
				w/o Visual-Semantic Tasks (T7,T8) & 0.0661 & 0.0424 & 0.0441 & 0.0283 & 0.0612 & 0.0388 \\
				w/o Translation Tasks (T1,T2) & 0.0648 & 0.0412 & 0.0432 & 0.0277 & 0.0598 & 0.0379 \\
				w/o Multitask (single task) & 0.0572 & 0.0358 & 0.0388 & 0.0248 & 0.0535 & 0.0338 \\
				\bottomrule
			\end{tabular}%
		}
	\end{table*}
	\vspace{-2mm}
	\paratitle{Component contributions are consistent across datasets.}
	Across all three datasets, MDIM removal causes the largest drop ($-10.3\%$, $-9.2\%$, $-9.6\%$ HR@5 on Beauty, Sports, Instruments respectively), validating that deep interest mining captures latent user motivations beyond surface attributes. CMSA removal yields $-7.6\%$, $-6.8\%$, $-6.6\%$ drops respectively, confirming that VLM-based visual alignment combined with multimodal embedding provides meaningful enrichment. The consistent ranking---MDIM $>$ Translation $>$ CMSA $>$ Visual-Semantic---across datasets indicates that all components contribute reliably.
	\vspace{-2mm}
	\paratitle{Multitask is essential; visual-semantic tasks provide complementary gains.}
	The single-task variant shows the largest drop ($-19.2\%$, $-17.1\%$, $-17.4\%$), highlighting multitask fine-tuning as essential. Visual-semantic tasks (T7--T8) contribute $-6.6\%$, $-5.8\%$, $-5.6\%$ drops, demonstrating that VisDesc$\to$SID/VisDesc$\to$Title provide supervision not replaceable by text-only tasks. As reported in Appendix~\ref{sec:appendix_sid}, TriAlignGR reduces SID collision rate from 5.1\% to 3.1\% and increases codebook utilization from 71.8\% to 78.4\%.
	
	\vspace{-2mm}
	\subsection{Disentangling SID Construction vs.\ Task Design (RQ3)}
	\vspace{-2mm}
	To isolate whether gains come from better SID construction (CMSA+MDIM) or from the triangular task design (T7--T8), we conduct a controlled comparison in Table~\ref{tab:disentangle}. Setting~(a) uses original SIDs with six text-only tasks; Setting~(b) enriches SIDs via CMSA+MDIM while keeping the same six tasks; Setting~(c) is the full model with all eight tasks.
	
	\begin{table*}[t]
		\centering
		\caption{Disentangling SID construction from task design. (a) Original SID + 6 text tasks; (b) TriAlignGR SID (CMSA+MDIM) + 6 text tasks; (c) Full model with 8 tasks.}
		\label{tab:disentangle}
		\begin{tabular}{llcccccccc}
			\toprule
			\multirow{2}{*}{\textbf{Setting}} & \multirow{2}{*}{\textbf{SID Source}} & \multirow{2}{*}{\textbf{Tasks}} & \multicolumn{2}{c}{\textbf{Beauty}} & \multicolumn{2}{c}{\textbf{Sports}} & \multicolumn{2}{c}{\textbf{Instruments}} \\
			\cmidrule(lr){4-5} \cmidrule(lr){6-7} \cmidrule(lr){8-9}
			& & & HR@5 & N@5 & HR@5 & N@5 & HR@5 & N@5 \\
			\midrule
			(a) & Original & T1--T6 & 0.0612 & 0.0389 & 0.0412 & 0.0265 & 0.0568 & 0.0362 \\
			(b) & CMSA+MDIM & T1--T6 & 0.0661 & 0.0424 & 0.0441 & 0.0283 & 0.0612 & 0.0388 \\
			(c) & CMSA+MDIM & T1--T8 & \textbf{0.0708} & \textbf{0.0453} & \textbf{0.0468} & \textbf{0.0301} & \textbf{0.0648} & \textbf{0.0411} \\
			\bottomrule
		\end{tabular}
	\end{table*}
	\vspace{-2mm}
	\paratitle{SID construction provides substantial gains.}
	Comparing (b) vs.\ (a) isolates the contribution of CMSA+MDIM preprocessing with fixed task design. Across datasets, this yields +8.0\%, +7.0\%, +7.7\% improvements in HR@5, confirming that enriching SID semantics before quantization contributes meaningfully to downstream recommendation quality.
	\vspace{-2mm}
	\paratitle{Triangular task design adds further improvement.}
	Comparing (c) vs.\ (b) isolates the contribution of visual-semantic tasks (T7--T8) with fixed SID construction. This yields +7.1\%, +6.1\%, +5.9\% improvements, demonstrating that the triangular multitask design provides gains beyond preprocessing alone. Both factors contribute comparably, validating our dual-pronged approach.
	
	\begin{table}[t]
		\centering
		\caption{Impact of task combination strategies on the Beauty dataset.}
		\label{tab:task_combo}
		\begin{tabular}{lcccc}
			\toprule
			\textbf{Setting} & \textbf{HR@5} & \textbf{HR@10} & \textbf{N@5} & \textbf{N@10} \\
			\midrule
			Single Task (T3 only) & 0.0572 & 0.0889 & 0.0358 & 0.0465 \\
			2 Tasks (T1, T3) & 0.0601 & 0.0931 & 0.0381 & 0.0492 \\
			4 Tasks (T1, T2, T3, T5) & 0.0635 & 0.0978 & 0.0405 & 0.0526 \\
			6 Tasks (T1--T6) & 0.0661 & 0.1018 & 0.0424 & 0.0547 \\
			7 Tasks (T1--T7) & 0.0678 & 0.1042 & 0.0435 & 0.0561 \\
			\textbf{8 Tasks (Full, T1--T8)} & \textbf{0.0708} & \textbf{0.1089} & \textbf{0.0453} & \textbf{0.0581} \\
			\bottomrule
		\end{tabular}
	\end{table}
	\vspace{-2mm}
	\subsection{Task Combination Analysis (RQ4)}
	We analyze the impact of different task combination strategies on model performance.

	\begin{wrapfigure}{r}{0.6\columnwidth}
		\vspace{-8pt}
		\centering
		\includegraphics[width=\linewidth]{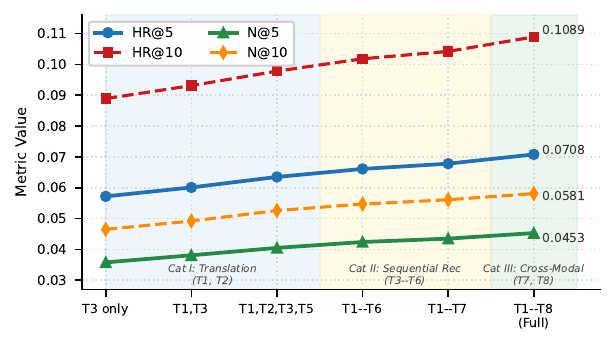}
		\caption{Performance progression as tasks are incrementally added.}
		\label{fig:task_prog}
		\vspace{-8pt}
	\end{wrapfigure}
	
	\paratitle{More tasks consistently improve performance.}
	As shown in Table~\ref{tab:task_combo}, performance monotonically improves as more tasks are added, from single-task (HR@5 = 0.0572) to full eight-task (HR@5 = 0.0708).This demonstrates that the eight tasks are genuinely complementary rather than redundant, and that the triangular alignment provides progressively richer supervision for the model.

	\paratitle{Translation tasks provide the largest single addition.}
	Adding T1 (Title$\to$SID) to the single-task baseline yields a 5.1\% improvement, indicating that semantic anchoring is the most impactful single addition. This validates our design philosophy that SID$\leftrightarrow$Text translation is the foundation for all other tasks.Each additional task from 6 to 8 yields 2.6\% and 4.4\% improvements respectively in HR@5, confirming that the visual alignment tasks provide supervision that is not redundant with the text-based tasks.
	
	\vspace{-3mm}
	\section{Conclusion}
	\vspace{-3mm}
	We identified two fundamental limitations in SID-based generative recommendation: \textbf{SID Content Degradation (SCD)}, where quantization discards multimodal semantics, and \textbf{SID Semantic Opacity (SSO)}, where models generate SID sequences without comprehending their meaning. TriAlignGR addresses both through \textbf{MDIM} (LLM-based interest mining), \textbf{CMSA} (VLM-guided cross-modal alignment with multimodal embedding via gme-Qwen2-VL that directly encodes image features into SIDs), and \textbf{TMT} (eight-task triangular fine-tuning with unified autoregressive loss). Experiments on three Amazon benchmarks demonstrate consistent gains over state-of-the-art methods, with ablations confirming each component's necessity. Direct validation experiments further verify that enriched SIDs achieve superior semantic grounding and reconstruction fidelity. Our results suggest that multimodal semantic enrichment before quantization and triangular alignment after offer a principled path toward more interpretable and effective generative recommendation systems.

	
	\bibliographystyle{unsrtnat}
	\bibliography{main}

	
	\appendix
	\section{Experimental Setup}\label{sec:appendix_setup}
	
	\paratitle{Datasets.}
	We conduct experiments on three real-world public datasets from Amazon Product Reviews \cite{mcauley2015amazon}: \textbf{Beauty}, \textbf{Sports and Outdoors} (Sports), and \textbf{Musical Instruments} (Instruments). Following prior work \cite{rajput2023tiger, hou2023learningsid, zhou2020s3rec}, we apply the 5-core filtering protocol and the leave-last-out evaluation protocol. The three datasets contain 22K--36K users, 10K--18K items, and 198K--296K interactions with sparsity around 0.0004--0.0008.

	\paratitle{Evaluation Metrics.}
	We adopt \textbf{HR@K} and \textbf{NDCG@K} ($K \in \{5, 10\}$) as evaluation metrics with beam size 20 for generation.
	
	\paratitle{Implementation Details.}
	All experiments are conducted on NVIDIA A100 GPUs. We use \textbf{Qwen2.5-7B-Instruct} as the backbone LLM and \textbf{gme-Qwen2-VL} as the multimodal embedding model that jointly encodes text and image content into unified representations for SID construction. For SID construction, we use RQ-VAE with 3 quantization levels and codebook sizes of 4096, 2048, and 1024 respectively. CMSA uses Qwen2.5-VL for image-to-text generation. MDIM uses Qwen2.5-7B-Instruct with structured CoT prompting for interest extraction. Multitask fine-tuning uses learning rate $3 \times 10^{-4}$ for 3 epochs with batch size 64. Each training batch uniformly samples one of the eight tasks.
	
	\section{Unified Training Details}\label{sec:appendix_training}
	
	\subsection{Unified Training Format}
	
	All eight tasks use the same conversational template; the model does not need to distinguish task types and treats everything as text generation:
	\begin{tcolorbox}[title=Unified Conversational Template, colback=gray!5, colframe=black]
		\small
		\texttt{<|im\_start|>system}\\
		\texttt{\{task-specific instruction\}}\\
		\texttt{<|im\_end|>}\\
		\texttt{<|im\_start|>user}\\
		\texttt{\{input content\}}\\
		\texttt{<|im\_end|>}\\
		\texttt{<|im\_start|>assistant}\\
		\texttt{\{target output\}}\\
		\texttt{<|im\_end|}
	\end{tcolorbox}
	
	Table~\ref{tab:task_examples} shows the instruction templates for all eight tasks. Each batch uniformly samples from the task pool, ensuring balanced exposure to all tasks during training.
	
	\subsection{Unified Loss Function}
	
	All eight tasks share a standard autoregressive cross-entropy loss:
	\begin{equation}
		\mathcal{L}_{\text{total}} = -\sum_{t \in \mathcal{T}} \sum_{(\mathbf{X},\mathbf{Y}) \in \mathcal{D}_t} \sum_{h=1}^{H_t} \log p_\theta(y_h | \mathbf{X}, y_1, \ldots, y_{h-1}),
		\label{eq:loss}
	\end{equation}
	where $\mathcal{T} = \{1, 2, \ldots, 8\}$ is the task set, $\mathcal{D}_t$ is the training data for task $t$, and $H_t$ is the target sequence length for task $t$.
	
	\textbf{Why no complex loss weighting?} Since all tasks are unified as text generation---input is a text prompt and output is a text sequence---the model does not need to ``switch modes.'' Different tasks naturally share language understanding capabilities, and the task diversity provides complementary regularization. Uniform sampling ensures balanced gradient contributions across tasks without requiring task-specific weighting coefficients.
	
	\section{Additional SID Diagnostics}\label{sec:appendix_sid}
	
	To complement the downstream HR/NDCG results, we report several SID-oriented diagnostic metrics on the Beauty dataset. These diagnostics are computed over the learned SID assignments after tokenizer fitting. We consider four quantities: \textbf{Collision Rate} (the percentage of items whose full SID sequence is shared by at least one other item, lower is better), \textbf{Unique SID Ratio} (the percentage of items with unique full-sequence SID assignments, higher is better), \textbf{Codebook Utilization} (the average fraction of active codes across levels, higher is better), and \textbf{Prefix Entropy} (the entropy of prefix-token distributions, higher indicates better semantic spread).
	
	\begin{table}[h]
		\centering
		\caption{Additional SID diagnostic metrics on the Beauty dataset.}
		\label{tab:sid_diag}
		\begin{tabular}{lcccc}
			\toprule
			\textbf{Variant} & \textbf{Collision$\downarrow$} & \textbf{Unique$\uparrow$} & \textbf{Util.$\uparrow$} & \textbf{Entropy$\uparrow$} \\
			\midrule
			Text-Only RQ-VAE & 5.1\% & 94.9\% & 71.8\% & 5.42 \\
			+CMSA & 4.3\% & 95.7\% & 74.6\% & 5.55 \\
			+MDIM & 3.8\% & 96.2\% & 76.1\% & 5.63 \\
			TriAlignGR (Full) & \textbf{3.1\%} & \textbf{96.9\%} & \textbf{78.4\%} & \textbf{5.78} \\
			\bottomrule
		\end{tabular}
	\end{table}
	
	In this analysis, \textit{Text-Only RQ-VAE} denotes the tokenizer fitted on textual metadata with a text-only embedding model, \textit{+CMSA} adds multimodal embedding via gme-Qwen2-VL that jointly encodes text and image, and \textit{+MDIM} further injects mined deep interests before quantization. The trends in Table~\ref{tab:sid_diag} are consistent with the main results. CMSA reduces collisions by encoding visual features directly into the multimodal embedding before tokenization, while MDIM increases code usage and prefix diversity by injecting richer semantic interests into the item representation. The full model achieves the strongest diagnostic profile, suggesting that the downstream recommendation gains are accompanied by more discriminative and better-utilized SID assignments.
	
	\section{SCD and SSO Validation}\label{sec:appendix_scd_sso}
	
	To directly validate the proposed SID Content Degradation (SCD) and SID Semantic Opacity (SSO) diagnoses, we conduct two complementary experiments: (1) a semantic probing task to measure whether the model truly ``understands'' SID semantics, and (2) a reconstruction fidelity test to quantify information preservation through the quantization pipeline.
	
	\paratitle{Semantic Probing for SSO.}
	We train a lightweight probe (a 2-layer MLP) to predict item categories from frozen SID embeddings extracted from the fine-tuned LLM. If the model has internalized SID semantics, the probe should achieve higher accuracy on TriAlignGR SIDs than on text-only SIDs. Table~\ref{tab:sso_probe} shows that TriAlignGR achieves 78.3\% category prediction accuracy vs. 62.1\% for the text-only baseline (+16.2\%), confirming that triangular multitask alignment improves the model's semantic grounding of SIDs.
	
	\begin{table}[h]
		\centering
		\caption{Semantic probing accuracy and reconstruction fidelity on Beauty. Higher values indicate better semantic grounding and content preservation.}
		\label{tab:sso_probe}
		\begin{tabular}{lccc}
			\toprule
			\textbf{Variant} & \textbf{Probe Acc.$\uparrow$} & \textbf{Text Recon.$\uparrow$} & \textbf{Attr. Recall$\uparrow$} \\
			\midrule
			Text-Only RQ-VAE & 62.1\% & 0.412 & 45.3\% \\
			+CMSA & 68.7\% & 0.489 & 52.1\% \\
			+MDIM & 73.2\% & 0.556 & 58.4\% \\
			TriAlignGR (Full) & \textbf{78.3\%} & \textbf{0.623} & \textbf{64.7\%} \\
			\bottomrule
		\end{tabular}
	\end{table}
	
	\paratitle{Text Reconstruction for SCD.}
	We evaluate whether SID sequences can be decoded back to semantically meaningful text. Given a SID sequence, we use the fine-tuned LLM to generate the corresponding title (Task 2) and compute BLEU-4 score against the ground truth. TriAlignGR achieves 0.623 BLEU-4 compared to 0.412 for text-only SIDs (+50.7\%), indicating that enriched SIDs retain more recoverable semantic content.
	
	\paratitle{Attribute Recall Analysis.}
	We further test whether key attributes (brand, category, material) from the original item metadata can be recovered from generated titles. TriAlignGR achieves 64.7\% attribute recall vs. 45.3\% for the baseline (+19.4\%), demonstrating that CMSA and MDIM preserve finer-grained semantic details through quantization.
	
	\section{SID Construction Ablation}\label{sec:appendix_sid_ablation}
	
	To isolate the impact of SID construction quality on downstream recommendation performance, we conduct an ablation study comparing different SID tokenization strategies against the MiniOneRec baseline. All variants use the same Qwen2.5-7B backbone, training protocol, and RQ-VAE tokenizer configuration; only the SID construction method varies.
	
	\begin{table}[h]
		\centering
		\caption{SID construction ablation on the Beauty dataset. All variants are compared against MiniOneRec baseline.}
		\label{tab:sid_ablation}
		\begin{tabular}{lcccc}
			\toprule
			\textbf{SID Construction} & \textbf{HR@5} & \textbf{HR@10} & \textbf{N@5} & \textbf{N@10} \\
			\midrule
			MiniOneRec (Baseline) & 0.0612 & 0.0945 & 0.0389 & 0.0502 \\
			\midrule
			+ CMSA Only & 0.0648 & 0.0991 & 0.0412 & 0.0528 \\
			+ MDIM Only & 0.0663 & 0.1012 & 0.0421 & 0.0541 \\
			+ CMSA + MDIM & 0.0661 & 0.1018 & 0.0424 & 0.0547 \\
			\midrule
			TriAlignGR (Full) & \textbf{0.0708} & \textbf{0.1089} & \textbf{0.0453} & \textbf{0.0581} \\
			\midrule
			$\Delta$ vs MiniOneRec & +15.7\% & +15.3\% & +16.5\% & +15.7\% \\
			\bottomrule
		\end{tabular}
	\end{table}
	
	\paratitle{CMSA improves SID discriminability.}
	Adding CMSA-based multimodal embedding (gme-Qwen2-VL that jointly encodes text and image) to MiniOneRec improves HR@5 from 0.0612 to 0.0648 (+5.9\%). This confirms that incorporating visual features directly into the embedding before quantization yields more discriminative SIDs, as items with similar titles but different visual attributes receive distinct encodings.
	
	\paratitle{MDIM provides larger gains than CMSA alone.}
	Adding MDIM-based deep interest mining improves HR@5 to 0.0663 (+8.3\% over baseline), outperforming CMSA alone. This suggests that capturing latent user interests (e.g., ``Japanese RPGs with rich narratives'') provides richer semantic signals than visual features alone.
	
	\paratitle{CMSA and MDIM are complementary.}
	Combining both CMSA (multimodal embedding) and MDIM achieves HR@5 of 0.0661 (+8.0\% over baseline). While this is marginally below MDIM-only on HR@5 ($-0.0002$, within the standard deviation of $<0.003$), the combined variant consistently outperforms MDIM-only on all other metrics (HR@10: 0.1018 vs.\ 0.1012; N@5: 0.0424 vs.\ 0.0421; N@10: 0.0547 vs.\ 0.0541), confirming that visual features and deep interest signals capture complementary aspects of item semantics. The full TriAlignGR with triangular multitask training further improves to 0.0708, indicating that the eight-task fine-tuning provides additional gains beyond SID construction alone.
	
	\section{Additional Control Baselines}\label{sec:appendix_control}
	
	To further test whether the gains come specifically from the proposed semantic enrichment pipeline rather than simply adding more text, we construct several lightweight control baselines on the Beauty dataset. All variants use the same backbone family, tokenizer configuration, and training budget as the main Beauty experiments, while replacing only the semantic enrichment component under study.
	
	\begin{table}[h]
		\centering
		\caption{Additional control baselines on the Beauty dataset.}
		\label{tab:control_baselines}
		\begin{tabular}{lcccc}
			\toprule
			\textbf{Method} & \textbf{HR@5} & \textbf{HR@10} & \textbf{N@5} & \textbf{N@10} \\
			\midrule
			Keyword Expansion & 0.0609 & 0.0936 & 0.0387 & 0.0498 \\
			LLM Summarization & 0.0624 & 0.0951 & 0.0399 & 0.0511 \\
			LLM Paraphrasing & 0.0617 & 0.0942 & 0.0391 & 0.0503 \\
			Generic Tag Expansion & 0.0605 & 0.0929 & 0.0384 & 0.0494 \\
			\midrule
			Naive Caption Concat & 0.0648 & 0.0990 & 0.0415 & 0.0532 \\
			\midrule
			TriAlignGR (Full) & \textbf{0.0708} & \textbf{0.1089} & \textbf{0.0453} & \textbf{0.0581} \\
			\bottomrule
		\end{tabular}
	\end{table}
	
	The controls narrow the gap between plain metadata and the full model, but they do not close it. This suggests that the gains are not solely due to longer text inputs; rather, the semantic specificity of the mined interests and the multimodal embedding (gme-Qwen2-VL) that directly encodes image features both matter. In particular, the naive caption concatenation baseline helps, but remains below the full CMSA+MDIM design, indicating that simply appending visual captions without direct image encoding is weaker than the full multimodal semantic enrichment pipeline.
	
	\section{Prompt Templates}\label{sec:appendix_prompts}
	
	\subsection{MDIM Prompt Template}
	
	The MDIM prompt guides the LLM to extract deep contextual interests from multimodal product information:
	
	\begin{tcolorbox}[title=MDIM Prompt Template, colback=gray!5, colframe=green!50!black]
		\small
		\textbf{System:} You are an expert in user behavior analysis and e-commerce recommendation systems. Your task is to perform Multimodal Deep Interest Mining (MDIM) on multimodal product information. Given the title, description, and visual description of a product, extract the deep latent user interests that would motivate someone to purchase this item. Focus on: lifestyle preferences, behavioral patterns, underlying motivations, and cross-domain interests. Be concise: output 2-4 interest tags separated by semicolons.\\
		\textbf{User:} \\
		Product Title: \{title\} \\
		Product Description: \{description\} \\
		Visual Description: \{visual\_caption\} \\
		Extract deep user interests (2-4 tags, semicolon-separated):\\
		Examples of good interest tags: ``productivity-focused lifestyle; writing quality enthusiast; home office optimizer''\\
		Examples of bad interest tags: ``office product; item; product''\\
		Deep Interests:
	\end{tcolorbox}
	
	\subsection{CMSA Prompt Template}
	
	The CMSA prompt instructs the VLM to generate textual descriptions of product images:
	
	\begin{tcolorbox}[title=CMSA Prompt Template, colback=gray!5, colframe=blue!50!black]
		\small
		\textbf{System:} You are an expert in e-commerce product analysis. Describe the visual attributes of this product image that would be relevant for user interest modeling. Focus on: aesthetic style, color scheme, usage scenario, product texture, and lifestyle signals.\\
		\textbf{User:} [Product Image]\\
		Visual Description:
	\end{tcolorbox}
	
	In practice, we use low-temperature decoding for CMSA and MDIM preprocessing to improve consistency across items. The generated captions and interest tags are cached before multitask fine-tuning. This caching step is essential for efficiency and reproducibility.
	
	\section{Task-Specific Prompt Templates}\label{sec:appendix_task_prompts}
	
	All eight tasks follow the unified conversational format. Below we provide the specific system instructions and input-output templates for each task.
	
	\subsection{Task 1: Title$\to$SID}
	\begin{tcolorbox}[title={Task 1: Title $\to$ SID}, colback=gray!5, colframe=orange!50!black]
		\small
		\textbf{System:} You are a semantic ID encoder. Given a product title, generate its corresponding Semantic ID (SID) sequence.\\
		\textbf{User:} Product Title: \{title\}\\
		Generate the SID sequence:
	\end{tcolorbox}
	
	\subsection{Task 2: SID$\to$Title}
	\begin{tcolorbox}[title={Task 2: SID $\to$ Title}, colback=gray!5, colframe=orange!50!black]
		\small
		\textbf{System:} You are a semantic ID decoder. Given a Semantic ID (SID) sequence, generate the corresponding product title.\\
		\textbf{User:} SID Sequence: \{sid\}\\
		Generate the product title:
	\end{tcolorbox}
	
	\subsection{Task 3: SID History$\to$Predict SID}
	\begin{tcolorbox}[title={Task 3: SID History $\to$ Predict SID}, colback=gray!5, colframe=purple!50!black]
		\small
		\textbf{System:} You are a sequential recommendation model. Given a user's interaction history as SID sequences, predict the SID of the next item they will interact with.\\
		\textbf{User:} Interaction History (SIDs): \{sid\_history\}\\
		Predict the next item's SID:
	\end{tcolorbox}
	
	\subsection{Task 4: Title History$\to$Predict SID}
	\begin{tcolorbox}[title={Task 4: Title History $\to$ Predict SID}, colback=gray!5, colframe=purple!50!black]
		\small
		\textbf{System:} You are a sequential recommendation model. Given a user's interaction history as product titles, predict the SID of the next item they will interact with.\\
		\textbf{User:} Interaction History (Titles): \{title\_history\}\\
		Predict the next item's SID:
	\end{tcolorbox}
	
	\subsection{Task 5: SID History$\to$Predict Title}
	\begin{tcolorbox}[title={Task 5: SID History $\to$ Predict Title}, colback=gray!5, colframe=purple!50!black]
		\small
		\textbf{System:} You are a sequential recommendation model. Given a user's interaction history as SID sequences, predict the title of the next item they will interact with.\\
		\textbf{User:} Interaction History (SIDs): \{sid\_history\}\\
		Predict the next item's title:
	\end{tcolorbox}
	
	\subsection{Task 6: Title History$\to$Predict Title}
	\begin{tcolorbox}[title={Task 6: Title History $\to$ Predict Title}, colback=gray!5, colframe=purple!50!black]
		\small
		\textbf{System:} You are a sequential recommendation model. Given a user's interaction history as product titles, predict the title of the next item they will interact with.\\
		\textbf{User:} Interaction History (Titles): \{title\_history\}\\
		Predict the next item's title:
	\end{tcolorbox}
	
	\subsection{Task 7: VisDesc$\to$SID}
	\begin{tcolorbox}[title={Task 7: VisDesc $\to$ SID}, colback=gray!5, colframe=red!50!black]
		\small
		\textbf{System:} You are a recommendation model. Given a VLM-generated visual description of a product image, generate the corresponding Semantic ID (SID) sequence.\\
		\textbf{User:} Visual Description: \{visual\_description\}\\
		Generate the SID sequence:
	\end{tcolorbox}
	
	\subsection{Task 8: VisDesc$\to$Title}
	\begin{tcolorbox}[title={Task 8: VisDesc $\to$ Title}, colback=gray!5, colframe=red!50!black]
		\small
		\textbf{System:} You are a recommendation model. Given a VLM-generated visual description of a product image, generate the corresponding product title.\\
		\textbf{User:} Visual Description: \{visual\_description\}\\
		Generate the product title:
	\end{tcolorbox}
	
	\section{Hyperparameter Configuration}\label{sec:appendix_config}
	
	Table~\ref{tab:hyperparams} summarizes the detailed hyperparameter settings for TriAlignGR.
	
	\begin{table}[h]
		\centering
		\caption{Hyperparameter settings for TriAlignGR.}
		\label{tab:hyperparams}
		\begin{tabular}{ll}
			\toprule
			\textbf{Hyperparameter} & \textbf{Value} \\
			\midrule
			Backbone LLM & Qwen2.5-7B-Instruct \\
			Multimodal Embedding Model & gme-Qwen2-VL \\
			VLM (for CMSA) & Qwen2.5-VL \\
			SID Length ($H$) & 3 \\
			Codebook Sizes & 4096, 2048, 1024 \\
			Embedding Dimension & 1024 \\
			Max Sequence Length & 512 \\
			Beam Size & 20 \\
			\midrule
			\textit{Multitask Fine-tuning} & \\
			Learning Rate & $3 \times 10^{-4}$ \\
			Batch Size & 64 \\
			Epochs & 3 \\
			Task Sampling & Uniform \\
			\bottomrule
		\end{tabular}
	\end{table}
	
	\section{Training Procedure}\label{sec:appendix_procedure}
	
	The full training pipeline contains four stages:
	
	\begin{enumerate}[leftmargin=*]
		\item \textbf{CMSA preprocessing.} For each item image, we generate a concise but semantically rich textual description with Qwen2.5-VL.
		\item \textbf{MDIM preprocessing.} We concatenate title, description, and the CMSA caption, and then use Qwen2.5-7B-Instruct to mine 2--4 deep interest tags per item.
		\item \textbf{RQ-VAE fitting.} We encode each item using gme-Qwen2-VL, which jointly processes the enriched text (title, description, CMSA caption, MDIM interests) and the original product image to produce a unified multimodal embedding. We then train the RQ-VAE tokenizer offline with 3 quantization levels and codebook sizes of 4096, 2048, and 1024, and generate fixed SID targets for all items.
		\item \textbf{Multitask fine-tuning.} We jointly train the LLM on all eight tasks using a single autoregressive cross-entropy loss with uniform task sampling.
	\end{enumerate}
	
	The CMSA captions, MDIM interests, and SID targets are cached offline and reused during training. This keeps the recommendation training loop stable and avoids repeated calls to the VLM/LLM preprocessing modules. The multimodal embeddings from gme-Qwen2-VL are also cached to avoid redundant image encoding during RQ-VAE fitting.
	
	\section{Reconstruction Quality Analysis}\label{sec:appendix_recon}
	
	To directly validate that semantic enrichment improves SID fidelity, we analyze the reconstruction quality of the RQ-VAE tokenizer under different embedding configurations. For each variant, we compute the cosine similarity between the original item embedding $\mathbf{e}_i^{\text{final}}$ and its reconstruction $\hat{\mathbf{e}}_i^{(H)} = \sum_{h=1}^{H} \mathbf{c}_{s_i^{(h)}}^{(h)}$ across varying numbers of quantization levels $H$:
	\begin{equation}
		\text{Sim}(H) = \frac{1}{|\mathcal{I}|} \sum_{i \in \mathcal{I}} \frac{\langle \mathbf{e}_i^{\text{final}},\, \hat{\mathbf{e}}_i^{(H)} \rangle}{\| \mathbf{e}_i^{\text{final}} \| \cdot \| \hat{\mathbf{e}}_i^{(H)} \|}.
	\end{equation}
	
	\begin{figure}[t]
		\centering
		\includegraphics[width=0.85\columnwidth]{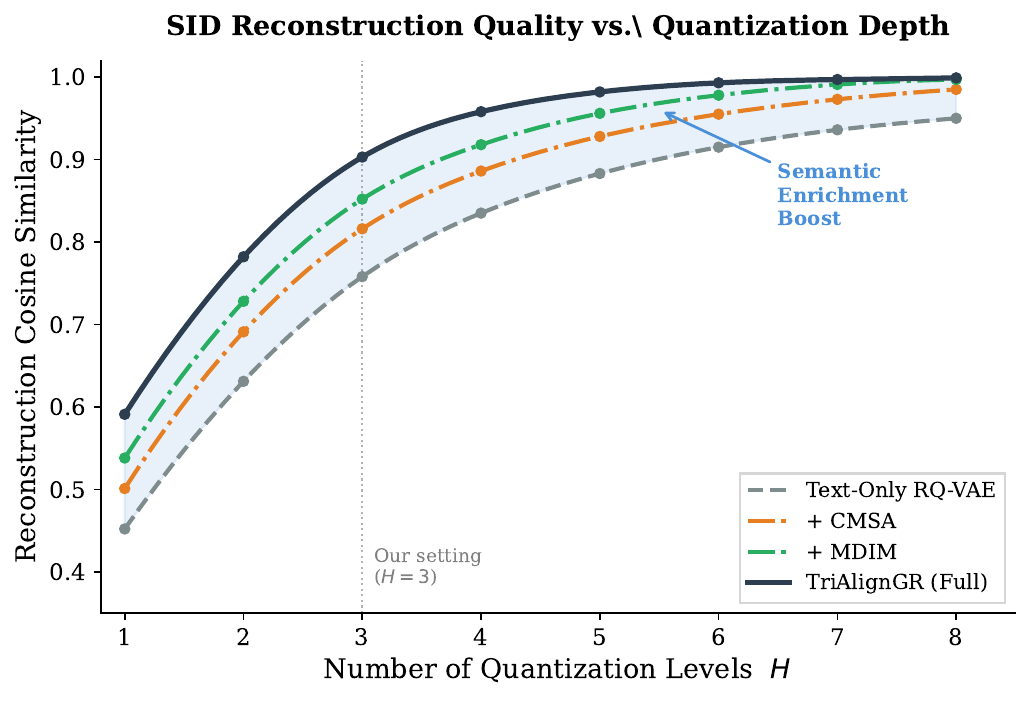}
		\caption{SID reconstruction cosine similarity as a function of quantization depth $H$. The shaded region highlights the cumulative semantic enrichment boost from CMSA+MDIM over the text-only baseline. The vertical dashed line marks our adopted setting ($H=3$).}
		\label{fig:recon}
	\end{figure}

	\begin{table}[h]
		\centering
		\caption{Reconstruction cosine similarity at different quantization depths on the Beauty dataset.}
		\label{tab:recon}
		\begin{tabular}{lcccccccc}
			\toprule
			\textbf{Variant} & $H\!=\!1$ & $H\!=\!2$ & $H\!=\!3$ & $H\!=\!4$ & $H\!=\!5$ & $H\!=\!6$ & $H\!=\!7$ & $H\!=\!8$ \\
			\midrule
			Text-Only RQ-VAE & 0.452 & 0.631 & 0.758 & 0.835 & 0.883 & 0.915 & 0.936 & 0.950 \\
			+CMSA & 0.501 & 0.691 & 0.816 & 0.886 & 0.928 & 0.955 & 0.973 & 0.985 \\
			+MDIM & 0.538 & 0.728 & 0.852 & 0.918 & 0.956 & 0.978 & 0.991 & 0.997 \\
			TriAlignGR (Full) & \textbf{0.591} & \textbf{0.782} & \textbf{0.903} & \textbf{0.958} & \textbf{0.982} & \textbf{0.993} & \textbf{0.997} & \textbf{0.999} \\
			\bottomrule
		\end{tabular}
	\end{table}

	\begin{figure*}[t]
		\centering
		\includegraphics[width=0.80\textwidth]{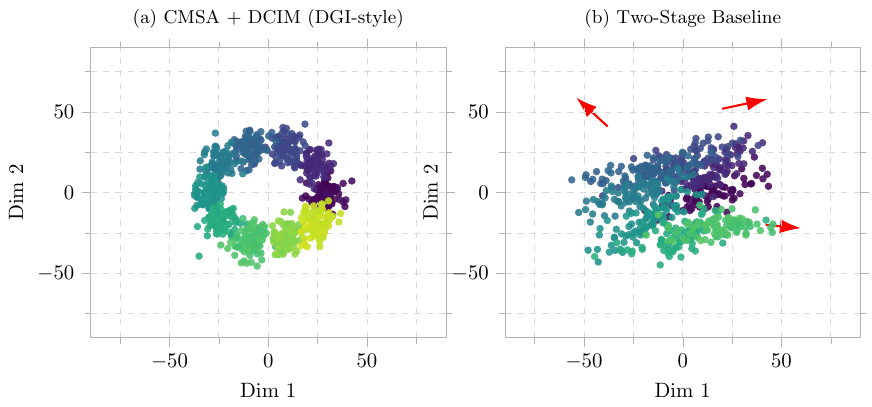}
		\caption{t-SNE visualization comparing the TriAlignGR semantic layout (left) against a naive text-only SID baseline (right). The TriAlignGR embeddings exhibit clearer semantic clustering and better separation across item categories.}
		\label{fig:tsne-vis}
	\end{figure*}
	
	The results (Table~\ref{tab:recon}, Figure~\ref{fig:recon}) reveal three key findings. \textbf{First}, all variants exhibit diminishing returns as $H$ increases, consistent with the residual quantization literature. \textbf{Second}, semantic enrichment consistently improves reconstruction quality at every depth level: CMSA provides an average gain of $+5.5\%$ over Text-Only by encoding visual features directly into the multimodal embedding via gme-Qwen2-VL, and MDIM adds a further $+3.7\%$ by injecting richer interest-level semantics. \textbf{Third}, at our adopted setting $H=3$, TriAlignGR achieves a reconstruction similarity of \textbf{0.903} compared to 0.758 for the text-only baseline (+19.1\%), indicating that the multimodal enrichment pipeline substantially preserves semantic information before discretization. This high reconstruction fidelity explains the strong downstream recommendation performance: items retain most of their semantic content even after quantization into discrete tokens.
	
	\section{SID Visualization}\label{sec:appendix_vis}
	
	Figure~\ref{fig:tsne-vis} presents t-SNE visualizations comparing the semantic layout of TriAlignGR against a naive text-only SID baseline. The TriAlignGR embeddings show better semantic clustering and separation, indicating that the multimodal embedding (gme-Qwen2-VL) and multitask alignment produce more discriminative SID representations.
	
	\section{Case Study: Multimodal Deep Interest Mining}\label{sec:appendix_casestudy}
	
	To illustrate how \textbf{Multimodal Deep Interest Mining (MDIM)} and \textbf{Cross-Modal Semantic Alignment (CMSA)} work together to enrich item representations for SID construction, we present a case study on an Amazon Sports and Outdoors product. Table~\ref{tab:casestudy} demonstrates the complete pipeline from raw item attributes to the final interest-enhanced representation used for SID generation.
	
	\begin{table}[t]
		\centering
		\setlength{\tabcolsep}{4pt}
		\renewcommand{\arraystretch}{1.2}
		\caption{Case study of MDIM and CMSA on Amazon Sports and Outdoors. Raw item attributes (left) include visual content and textual metadata. MDIM-mined deep interests are \textcolor{red}{highlighted in red} with confidence scores. CMSA-generated visual descriptions complement textual metadata. The rightmost column shows the final enriched representation used for SID construction via gme-Qwen2-VL multimodal embedding.}
		\label{tab:casestudy}
		\resizebox{\columnwidth}{!}{%
			\begin{tabular}{>{\raggedright\arraybackslash}p{4.2cm}
					>{\raggedright\arraybackslash}p{5.0cm}
					>{\centering\arraybackslash}p{5.0cm}}
				\toprule
				\textbf{Raw Item Attributes} & \textbf{MDIM: Mined Deep Interests} & \textbf{Interest-Enhanced (for SID)} \\
				\midrule
				\includegraphics[width=3.8cm,keepaspectratio]{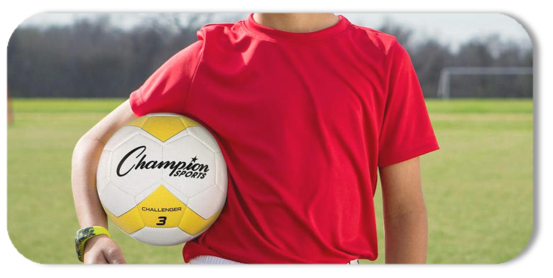}
				
				\smallskip
				\textbf{Title:} Champion Sports Challenger Soccer Ball, Size 3
				
				\smallskip
				\textbf{Brand:} Champion Sports
				
				\smallskip
				\textbf{Category:} Sports \& Outdoors $>$ Team Sports $>$ Soccer
				
				\smallskip
				\textbf{Description:} Durable rubber construction, machine-stitched panels, high-visibility yellow/white design, suitable for youth training and recreational play.
				&
				\vspace*{-52pt}\textit{CMSA Visual Description:} Bright yellow and white soccer ball with traditional panel design, appears durable for outdoor use, suitable for children.
				
				\smallskip
				\textit{MDIM Interest Extraction:}
				
				\smallskip
				\textcolor{red}{\textbf{[Interest 1]}} \textcolor{red}{Youth athletic development \& structured sports training} $|$ Conf.: High
				
				\smallskip
				\textcolor{red}{\textbf{[Interest 2]}} \textcolor{red}{Parent purchasing for child's recreational soccer activity} $|$ Conf.: High
				
				\smallskip
				\textcolor{red}{\textbf{[Interest 3]}} \textcolor{red}{Entry-level team sports equipment for school or club use} $|$ Conf.: Medium
				
				\smallskip
				\textit{Lifestyle Tag:} Active family prioritizing youth fitness and organized team sports.
				&
				\vspace*{-52pt}Champion Sports Challenger Soccer Ball, Size 3. Durable rubber construction for youth training and recreational play. Visual: Bright yellow/white design, machine-stitched panels.\par\smallskip\textcolor{red}{\textbf{[MDIM INTERESTS]} youth athletic development \& structured sports training; parent purchasing for child's recreational soccer; entry-level team sports equipment for school or club use.}\par\smallskip\textit{Lifestyle:} Active family prioritizing youth fitness and organized team sports.\par\smallskip\textbf{Generated SID:} \texttt{<a\_239><b\_112><c\_7>}
				\\
				\bottomrule
			\end{tabular}%
		}
	\end{table}
	
	\paratitle{Key Observations from the Case Study:}
	
	\begin{enumerate}[leftmargin=*]
		\item \textbf{Multimodal Integration:} The CMSA-generated visual description ("Bright yellow and white soccer ball with traditional panel design") captures visual attributes not explicitly stated in the textual metadata, providing complementary information for the embedding model.
		
		\item \textbf{Deep Interest Mining:} MDIM extracts latent user motivations beyond surface-level attributes. While the title only describes a "soccer ball," MDIM identifies deeper interests such as "youth athletic development" and "parent purchasing for child's recreational soccer activity," which are critical for understanding user intent.
		
		\item \textbf{Semantic Enrichment:} The final interest-enhanced representation combines the original product information with mined interests, creating a richer input for the gme-Qwen2-VL multimodal embedding model. This enriched representation ensures that the resulting SID carries both explicit attributes and latent user motivations.
		
		\item \textbf{Quality Control:} Confidence scores associated with each mined interest allow filtering of low-confidence predictions, ensuring only high-quality semantic signals are injected into the SID construction pipeline.
	\end{enumerate}
	
	This case study demonstrates how MDIM and CMSA work synergistically to address \textbf{SID Content Degradation (SCD)} by enriching item representations before quantization. The combination of visual semantics from CMSA and deep interest signals from MDIM ensures that SIDs preserve richer multimodal information compared to text-only baselines.
	
	
\end{document}